\documentclass[%
 jor,
 amsmath,amssymb,
 reprint,%
]{revtex4-2}
\usepackage{graphicx,mhchem}
\usepackage{dcolumn}
\usepackage{bm}
\usepackage{siunitx}
\usepackage{color}

\newcommand{\red}{\textcolor{red}}
\usepackage{afterpage}
\begin{document}


\title[Si solar cell]{Bessel-beam direct-write of the etch-mask in a nano-film of alumina for high-efficiency Si solar cells}

\author{
Tomas Katkus$^1$, Soon Hock Ng$^1$, Haoran Mu$^1$, Nguyen Hoai An Le$^1$, Dominyka Stonyt\.{e}$^2$,  Zahra Khajehsaeidimahabadi$^1$, Gediminas Seniutinas$^1$, Justas Baltrukonis$^3$, Orestas Ul\v{c}inas$^3$, Mindaugas Mikutis$^3$, Vytautas Sabonis$^3$, Yoshiaki Nishijima$^4$,  Michael Rien\"{a}cker$^5$,  Jan Kr\"{u}gener$^6$, Robby Peibst$^{5,6}$, Sajeev John$^7$, Saulius Juodkazis$^{1,2,8}$
}
\affiliation{$^1$ Optical Sciences Centre, ARC Training Centre in Surface Engineering for Advanced Materials (SEAM), Swinburne University of Technology, Hawthorn, Victoria 3122, Australia}
\affiliation{$^2$ Laser Research Center, Physics Faculty, Vilnius University, Saul\.{e}tekio Ave. 10, 10223 Vilnius, Lithuania}
\affiliation{$^3$ Altechna R\&D, 
Mokslinink\c{u} str. 6A, 08412 Vilnius, Lithuania}
\affiliation{$^4$Department of Electrical and Computer Engineering, Graduate School of Engineering, Yokohama National University, 79-5 Tokiwadai, Hodogaya-ku, Yokohama, 240-8501, Japan}
\affiliation{$^5$ Institut f\"{u}r Solarenergieforschung Hameln (ISFH), Am Ohrberg 1, 31860 Emmerthal, Germany}
\affiliation{$^6$ Institute of Electronic Materials and Devices, Leibniz Universit\"{a}t Hannover, Schneiderberg 32, 30167 Hannover, Germany}
\affiliation{$^7$Department of Physics, University of Toronto, 60 St. George Street, Toronto, ON, M5S 1A7, Canada }
\affiliation{$^8$ WRH Program International Research Frontiers Initiative (IRFI) Tokyo Institute of Technology, Nagatsuta-cho, Midori-ku, Yokohama, Kanagawa 226-8503 Japan}
%

\date{\today}

\begin{abstract}
Large surface area applications such as high-efficiency $ > 26\%$ solar cells require surface patterning with 1-10~$\mu$m periodic patterns at high fidelity over 1-10~cm$^2$ areas (before up scaling to 1~m$^2$) to perform at, or exceed, the Lambertian (ray optics) limit of light trapping. Here we show a pathway to high-resolution sub-1~$\mu$m etch mask patterning by ablation using direct femtosecond laser writing performed at room conditions (without the need for a vacuum-based lithography approach). A Bessel beam was used to alleviate the required high surface tracking tolerance for ablation of 0.3-0.8~$\mu$m diameter holes in $\sim 40$~nm alumina \ce{Al2O3}-mask at high writing speed, 7.5~cm/s; a patterning rate 1~cm$^2$ per 20~min. Plasma etching protocol was optimised for a zero-mesa formation of photonic crystal (PhC) trapping structures and smooth surfaces at the nanoscale level. Scaling up in area and throughput of the demonstrated approach is outlined.   
\end{abstract}

\keywords{Si solar cell, Lambertian limit, high-efficiency solar-to-electrical energy conversion, Bessel beam}
\maketitle

\section{\label{Intro}Introduction}


Si solar cells are the most developed practical solution for harvesting solar energy and are widely available at a $>23\%$ efficiency of solar-to-electrical power 
conversion at the commercial photo-voltaic (PV) module level for a household and industrial usage as install-ready $\sim 2$~m$^2$ panels~\cite{report,Saga,C5EE03380B}. The technology of Si solar cells is well-matured and produces energy at a cost lower than that traditionally available from national power grids distributing electricity from power stations (coal, gas, nuclear)~\cite{Ram2018a}. Recently, 
the world record performance has been improved 
from the previously achieved efficiency of Si solar cells 26.81\%~\cite{Lin2023} and stands now at 27.1\% 
measured by the National Renewable Energy Laboratory (NREL)~\cite{nrel}. An improvement by 1\% in the efficiency took $\sim 25$ years from M. Green's established 25\% benchmark for the monocrystalline Si demonstrated in 1999~\cite{Kaneka}. None of the modern solar cells are performing at the Lambertian (ray optics) limit of light trapping $4n^2$, where $n$ is the refractive index of solar cell material~\cite{limit}. The practical limit is predicted at 29.1\%, which is lower than the theoretical Shockley–Queisser 32\% boundary for Si at one Sun~\cite{Kaneka}. 

Random micro-textured surfaces are used to reduce reflectivity and enhance light trapping, however, they fundamentally cannot perform above the Lambertian limit, which requires harnessing the interference, slow light, and normal-to-parallel redirection of incident light. By harnessing those wave properties provide an enhanced absorbance at strongly localized resonances, especially at the energies close to the bandgap~\cite{Aydin2011}. The first experimental demonstration of light absorption above the Lambertian limit in Si slab was made using PhC structures with etch masks patterned by electron beam lithography (EBL) with subsequent wet etching in KOH and plasma~\cite{SJohn2}. The possibility to define etch mask by stepper projection lithography and direct laser write (DLW), which both have the capability of larger area patterning, was demonstrated for Si wafer~\cite{22oea210086}. The same PhC pattern, which performed the above Lambertian light trapping, was made on actual Si solar cells 
and compared to a neighboring central cell 
of the same geometry and structure on the same wafer but without PhC patterns (7 solar cells in total per wafer)~\cite{SiliconPV23}. This first demonstration of PhC light trapping on a Si solar cell showed 23.1\% efficiency on the PhC textured region (independently verified by ISFH CalTeC) as compared to 21.1\% for the neighboring central cell on the same wafer.  Another active trend in solar cells is making them thin, flexible and low-weight~\cite{oyo} including use of 2D materials~\cite{2D}. PhC light trapping can be used to achieve 95\% absorption of light within 10~$\mu$m thick Si solar cell to expand their application potential~\cite{SJohn1}. Flexible Si solar cells of 54~$\mu$m thickness showed the highest power-to-weight ratio of 1.9~W/g for $\sim 274$~cm$^2$ area~\cite{flexi}; the light-to-electrical power converison was 26.06\%.      
\begin{figure}[tb]
    \centering\includegraphics[width=8.5cm]{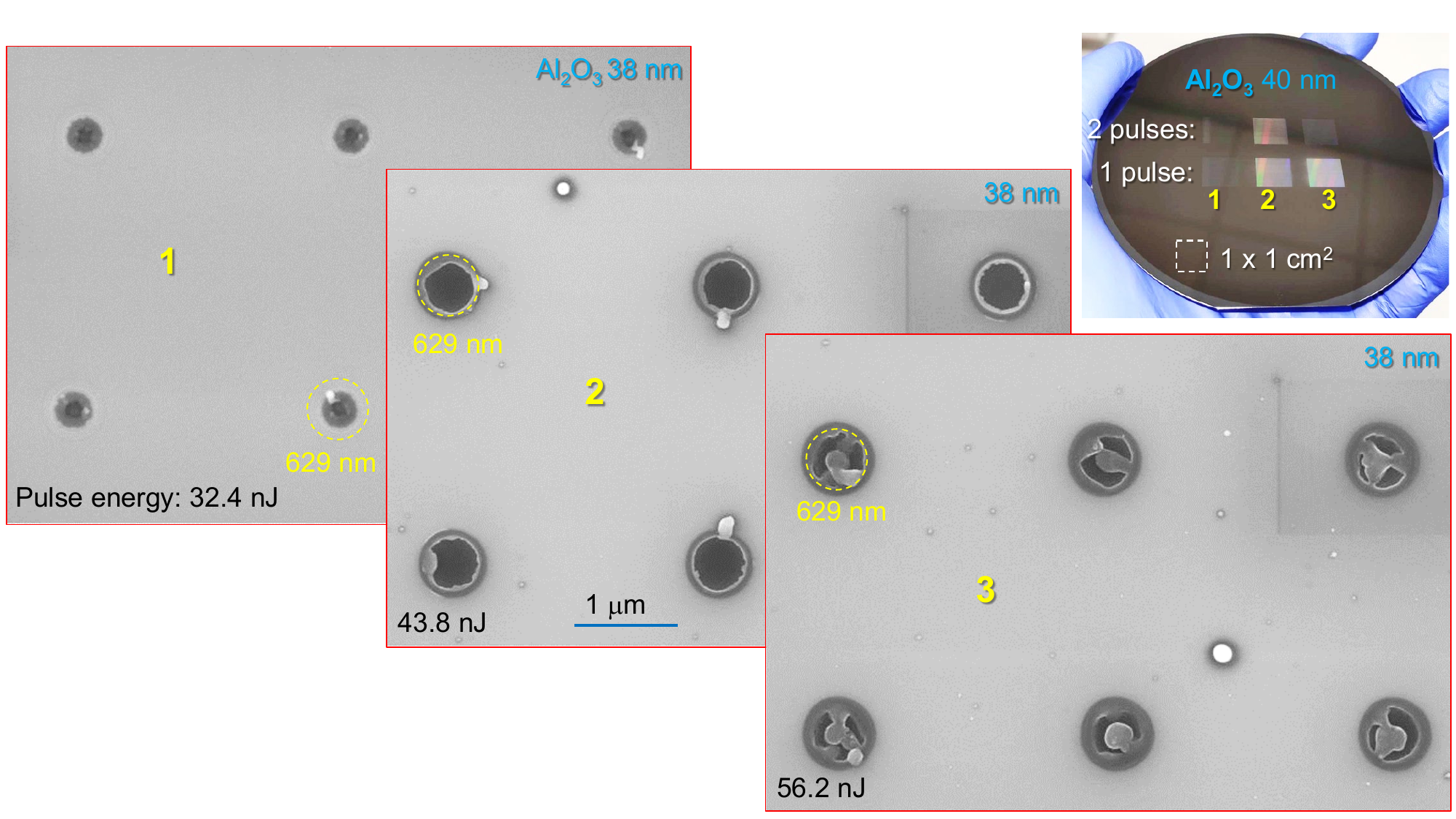}      \caption{\label{f-dumy} SEM images of 38-nm-thick ($n+i\kappa = 1.595 + i\times 0.01054$) \ce{Al2O3} mask ablated by single pulses 515~nm/170~fs (Carbide, Light Conversion) using an axicon lens, which formed the Bessel beam. Pulse energies were $E_p\approx 32, 44, 56$~nJ (marked on images). Photo image of a Si wafer with $\sim$40-nm-thick mask ablated in one and two pulses per site modes. The circle marker of 629~nm corresponds to a barely recognizable modification at the smallest pulse energy. This diameter is smaller than the $1/e^2$-intensity cross-section of the central core of the Bessel beam which was $\sim 880~$nm (the first minimum of the Bessel intensity envelope).   
}
\end{figure}
\begin{figure*}[t!]
    \centering\includegraphics[width=18cm]{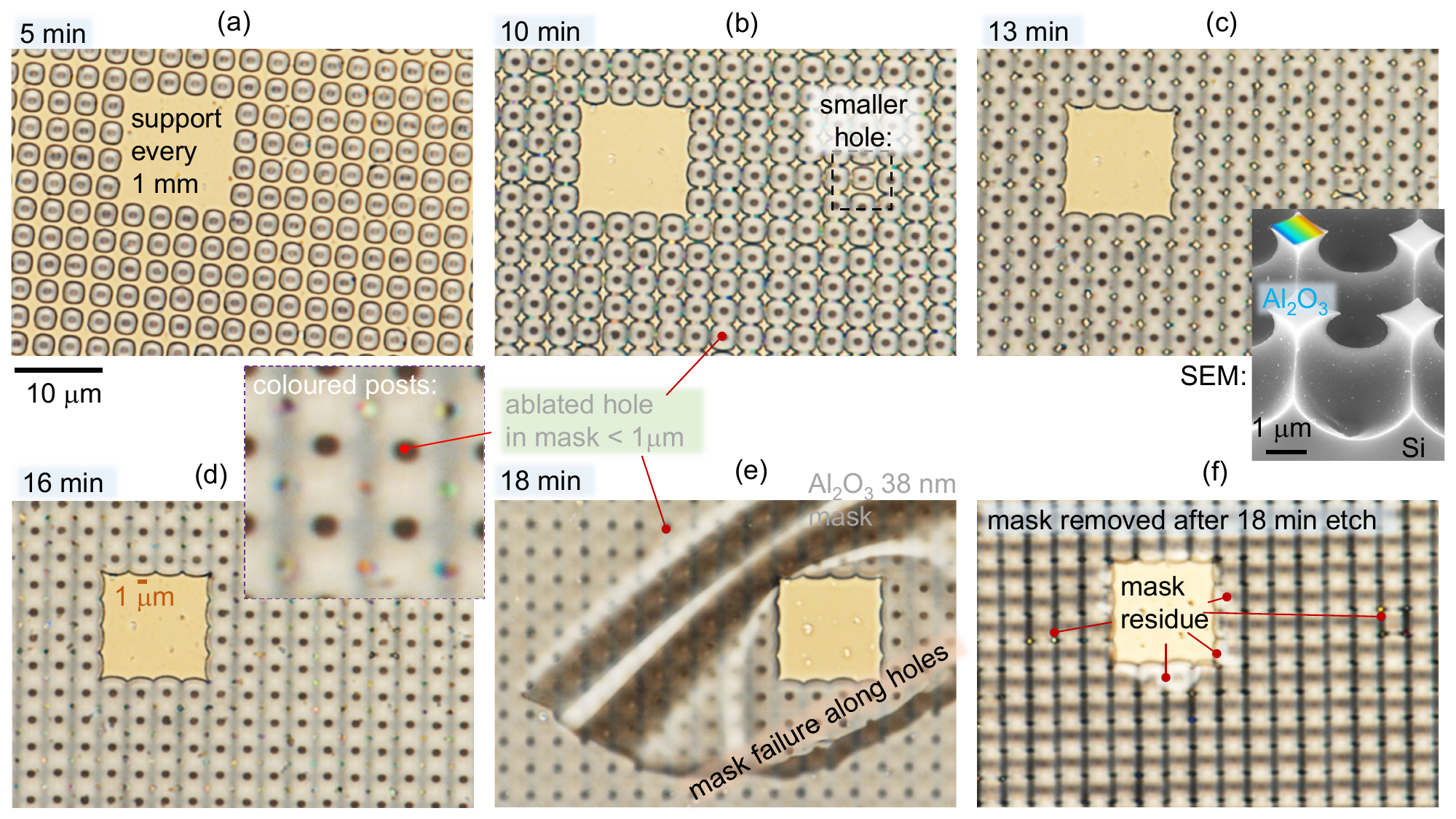}      \caption{\label{f-fast} Optical microscopy observation of etch time progression  of Photonic Crystal (PhC) light trapping surface on p-type Si (783 wafer); microscope Nikon Optiphot-pol. with an objective lens of numerical aperture $NA = 0.9$; see top-markers in (a-f)  for the 5-18~min span). Reactive ion etching (RIE; Samco RIE-101iPH) at high etch rate of isotropic mode at 200~W of Inductively Coupled Plasma (ICP) power and at 0~W Bias (directional etch). Chemistry: 50/5~cm$^3$/min (or sccm) of \ce{SF6}/\ce{O2}; process pressure 2~Pa, He cooling pressure 1~Pa. 
    The inset in (c) is a Scanning Electron Microscopy (SEM) image of etched PhC at 50/0~sccm of \ce{SF6}/\ce{O2} after 15~min. It illustrates the mesa-structures and \ce{Al2O3} mask adhered at the corner positions of the square pits (a rainbow-marker). When those corner-structures become smaller than $\sim 1~\mu$m in size (close to the diffraction limit of optical observation $0.61\lambda/NA$ at central visible wavelength $\lambda = 530$~nm-green), they appear colorful (see inset in (d)). This condition indicates an optimal time for the smallest mesa segments. Longer etch cause failure of the alumina mask shown in (e). Removal of 38-nm-thick alumina mask was made in ultra-sonic bath in millipore water. Mask writing was made with the Bessel beam. 
}
\end{figure*}
\begin{figure*}[tb]
    \centering\includegraphics[width=18cm]{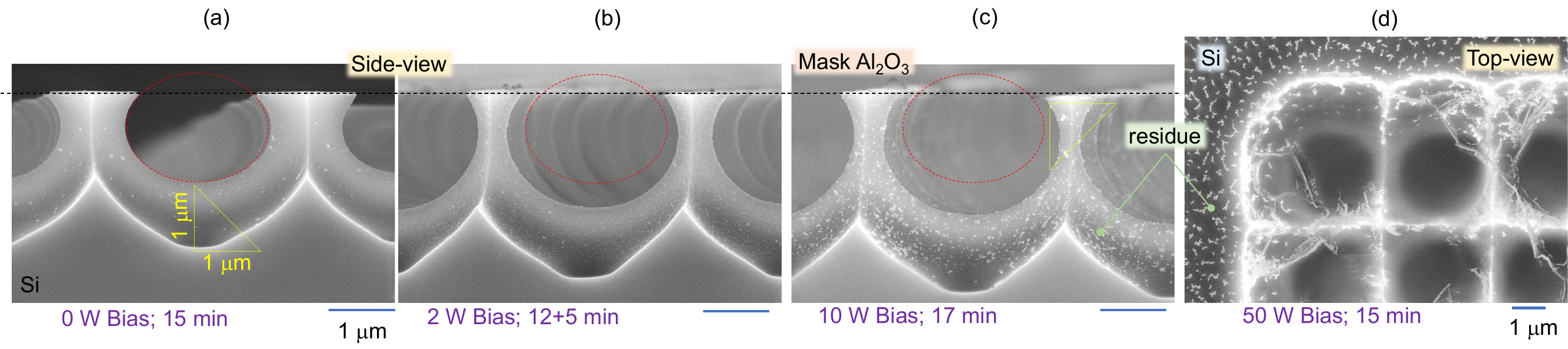}      \caption{\label{f-bias} SEM images of etch cross sections at different Bias powers: 0~W (a), 2~W (b), 10~W (c) and 50~W (top-view in (d)). Etch time is shown at the bottom of panels; when repeated etching was applied, times were shown with a summation marker ``+'' (b). RIE was carried out at 2 Pa process pressure, 200~W ICP power at 50/0~sccm \ce{SF6}/\ce{O2} gas mixture. For comparison of evolution of etched pattern oval and right-triangular markers of the same size are shown in (a-c). Thickness of \ce{Al2O3} mask was 28.4~nm; mask writing was with Bessel beam. Residue of oxidised debris increased with larger bias powers and after repeated etching runs.
}
\end{figure*}
\begin{figure*}[tb]
    \centering\includegraphics[width=13.5cm]{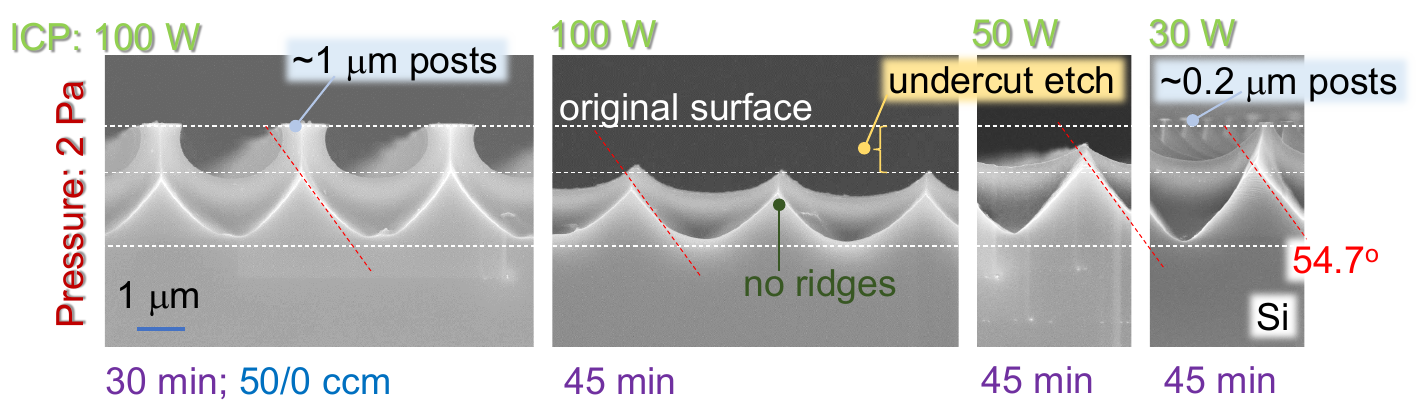}      \caption{\label{f-under} Evolution of under-etching at decreasing ICP power from 100~W to 30~W (the final optimised condition) at 0~W Bias; \ce{SF6}/\ce{O2} 50/0~sccm, pressure of 2~Pa. The under-etching removes the post regions. The minimum size of the posts also corresponded at the largest depth of the etched pattern (appeared colorful under optical observation; Fig.~\ref{f-blue}). The timing of etching is the main parameter for the optimal conditions of PhC formation. Thickness of \ce{Al2O3} mask was 38~nm; mask writing was with Bessel beam. The angle of $54.7^\circ$ between top surface (100) plane and (111) plane is shown by dashed-inclined lines.
}
\end{figure*}
\begin{figure}[tb]
    \centering\includegraphics[width=8.5cm]{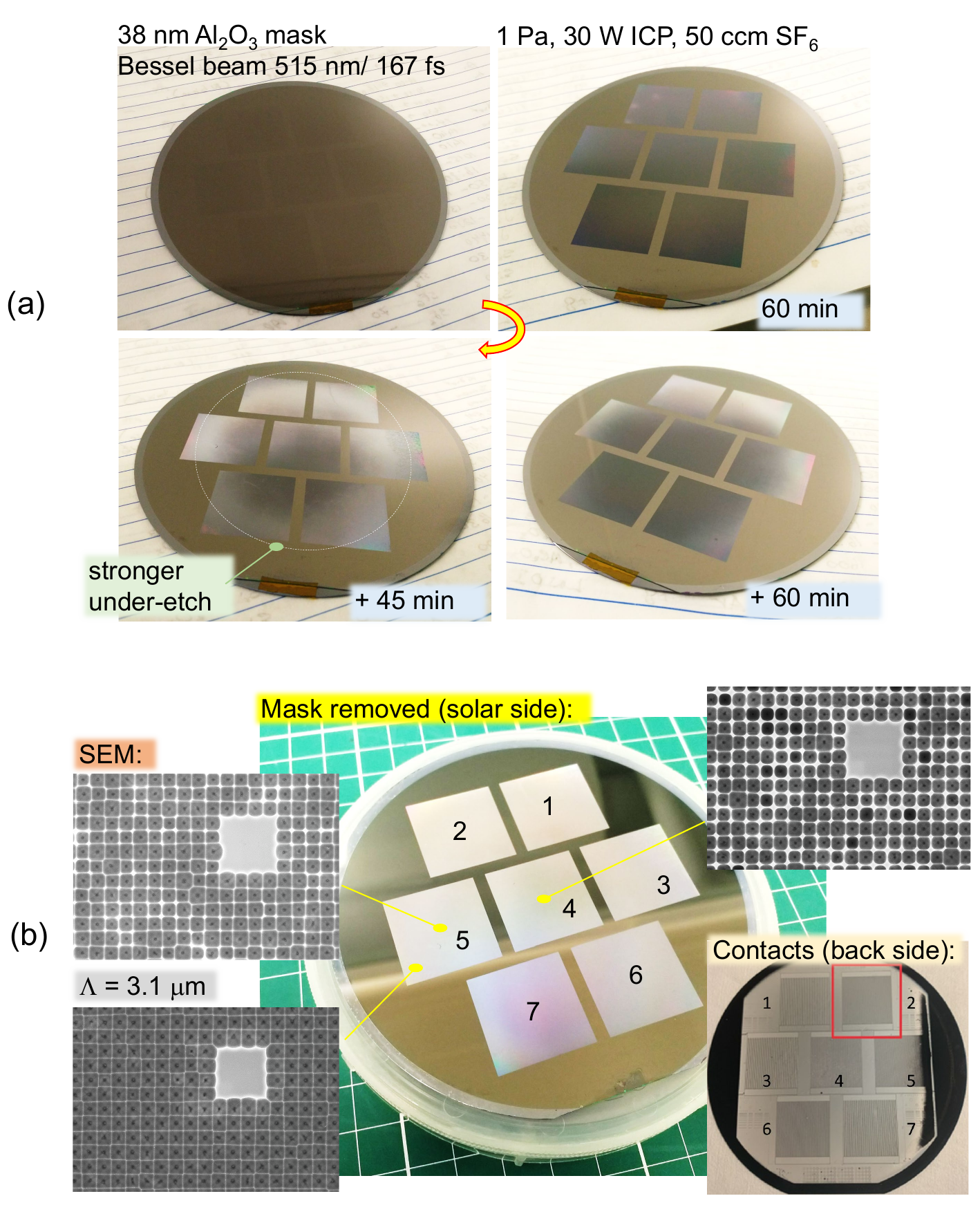}      \caption{\label{f-after} PhC light trapping for Polysilicon on Oxide  Interdigitated Back Contacted - POLO$^2$-IBC cell. 
    The thickness of Si solar cell 190~$\mu$m.  (a) The mask (38-nm-thick \ce{Al2O3}) was patterned by fs-laser direct write with Bessel beam 515~nm/167~fs, pulse energy $E_p = 43.8$~nJ over $2.2\times 2.2$~cm$^2$ areas on the front-side above of an integrated Si cells with back-side contacts. Plasma etch and PhC pattern evolution were observed at different paused times during etching. The optimised protocol was used: 1~Pa process pressure, 30~W ICP power, \ce{SF6}/\ce{O2} 50/0~ccm flow rate.   (b)
    Center-image is the final PhC trapping surface after mask removal in an IPA ultrasonic bath. 
     Insets show SEM images of the final PhC light trapping surface at different locations. There was a slight difference in terms of the under-etch conditions across the wafer recognizable in optical (a) and SEM (b) images. The right-bottom inset image shows the back-side contacts. 
}
\end{figure}
\begin{figure}[b!]
    \centering\includegraphics[width=8.5cm]{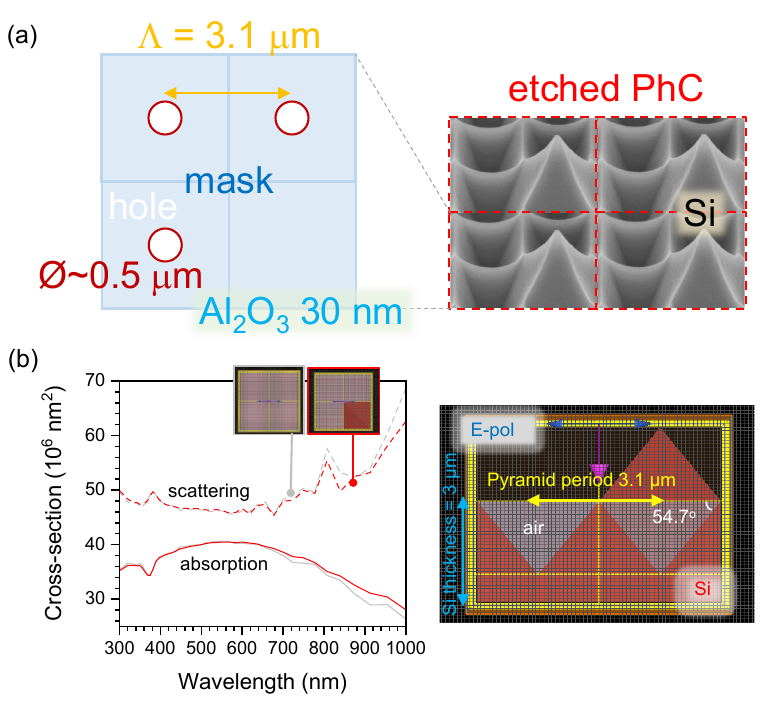}      \caption{\label{f-dream} (a) Concept for a better mechanical support of the nano-film mask over large areas for a deeper etch of PhC patterns:  laser inscribed mask in $\sim 30$-nm-thick \ce{Al2O3} with periodic micro-posts (not ablated sites). (b) Finite difference time domain (FDTD) calculations of absorption and scattering cross sections for the pattern in (a) and same geometry only with four pyramidal pits. 
    Periodic boundary conditions are applied for the xy-plane (lateral dimension) and perfectly matching layers in the longitudinal direction (along light propagation). The thickness of the cell is 3~$\mu$m; Si permittivity $(n+i\kappa)^2$ is from the Lumerical, Ansys database. Geometrical cross-section (footprint) of the pyramid (or inverted pyramid) is $9.61\times 10^6$~nm$^2$.
}
\end{figure}
\begin{figure*}[t!]
    \centering\includegraphics[width=18cm]{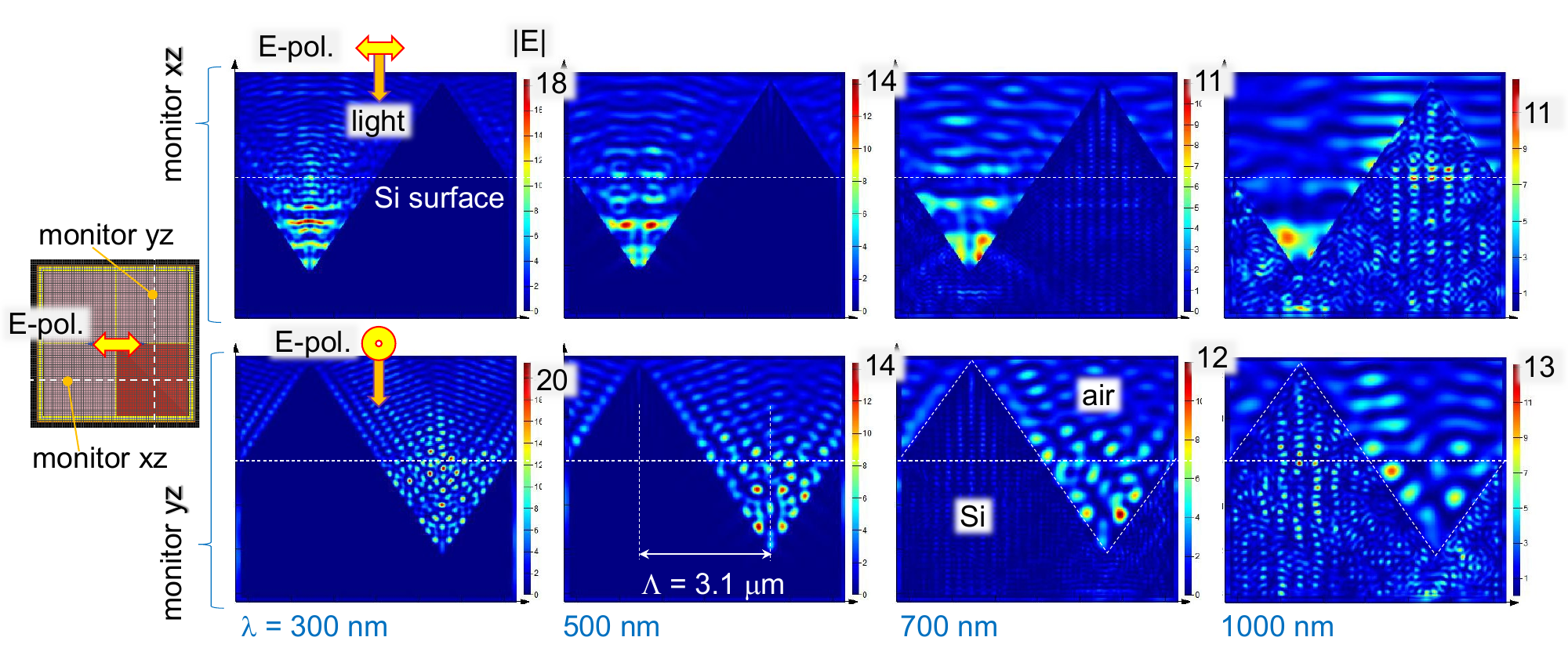}      \caption{\label{f-fdtd} 
    Light intensity enhancement at two cross-sections (monitors) at difference wavelengths simulated by FDTD with Periodic boundary conditions applied for the lateral dimension and perfectly matching layers in the longitudinal direction (along light propagation); same geometry as in Fig.~\ref{f-dream}. The thickness of the cell is 3~$\mu$m; Si permittivity $(n+i\kappa)^2$ is from the Lumerical, Ansys database. Incident light intensity $|E|^2=1$.
}
\end{figure*}

Here we demonstrate the fabrication of PhC pattern for light trapping on the face surface of actual Si solar cells. Dielectric \ce{Al2O3} mask of tens-of-nm thickness was laser ablated by ultra-short laser pulses (515~nm/167~fs) using an axicon lens for the formation of the Bessel beam~\cite{Kishan}. Ablated holes in the mask had a diameter smaller than the diffraction limit at the focusing conditions used due to the threshold effect, i.e., the ablation threshold of the mask was exceeded only in the very center of the focal spot. The dry plasma etching protocol was optimized for a nano-smooth surface of PhC pattern with minimal ridges between inverted pyramidal pits of PhC, without apparent debris and oxidised nanoscale regions. Optimization of plasma etching over a wide parameter space of chemistry used, process pressure, and flow rates of gasses is presented. The PhC light trapping structures were fabricated on  Polysilicon on Oxide  Interdigitated Back Contacted - POLO$^2$-IBC cell; such cells have the potential to perform at $>26\%$ efficiency of solar-to-electrical power conversion~\cite{HAASE2018184}.    

\section{Methods and materials}

The thickness of the electron (e)-beam (JKLesker AXXIS) coated \ce{Al2O3} was determined by ellipsometry (Woolam, VASE). Masks of 30~nm and 40~nm were investigated for optimisation of the laser ablation and plasma etch protocols. The refractive index $n+i\kappa \approx 1.595 + i0.011$ of \ce{Al2O3} was determined by ellipsometry. Thinner 20-nm-masks showed up to a few percent reduction in $\kappa$ while $n$ was proportionally larger.  

Si wafers (783)  were p-type (Boron doped) with resistivity $1-10~\Omega\cdot$cm and were close to that of the actual POLO$^2$-IBC p-type solar cells. The wafer's cut edge (see inset in Fig.~\ref{f-dumy}) was used to align the PhC mask pattern. The alignment is critical for plasma etching while low ICP power is used and the etching result resembles structures, similar to those obtained by wet etching in KOH with a distinct pattern of inverted pyramids. Imperfect mask alignment causes defects in mesa ridges and corner posts, which should be minimized for the best-performing light trapping. 

The axicon used in this study was designed for $\lambda = 515$~nm wavelength and it was fs-laser inscribed inside the bulk of silica glass. The form-birefringence of nano-gratings was used to engineer the refractive index of concentric grating for the required phase retardance. The cone angle of the flat axicon corresponded to 178$^\circ$. A $f = 400$~mm lens combined with  $ \times 20 $ magnification, $NA = 0.4$ objective lens was used to scale down the Bessel beam using an optical 4F relay scheme. 
The first minimum of the Bessel intensity profile is at $d_{min}^{B} = 0.383\lambda/\sin\gamma$, where $\lambda = 515$~nm is the laser wavelength, $\gamma = 18.46^\circ$ is the cone angle with respect to the optical axis along beam propagation; $d_{min}^{B} = \frac{4.816}{k_\perp}$, where $k_\perp = k\sin\gamma$ is the wavevector component perpendicular to the propagation ($k=2\pi/\lambda$). 
The final cone angle on the sample was $\gamma = 18.46^\circ$, 
the diameter of $d_{min}^B = 1.24~\mu$m at the first minimum or 0.88~$\mu$m at $1/e^2$-intensity level. The length of the non-diffractig Bessel beam zone was $\sim 400~\mu$m. 

The central spot as well as every ring of the Bessel beam in the focal region carries the same amount of energy. The average intensity and fluence per pulse of the optical $E_p = 44$~nJ energy used in mask patterning is estimated considering the central spot of the diameter equal to the first minimum $d_{min}^{B}$: $I_p\equiv F_p/t_p = 21.76$~TW/cm$^2$ and $F_p = E_p/(\pi [d_{min}^B/2]^2) =3.63$~J/cm$^2$, respectively, for the $t_p = 167$~fs pulse ($ct_p = 50~\mu$m axial extent). 
This calculation is carried out the same way as for the Gaussian pulse and can be used for protocol development. For the actual comparison of fluences between Bessel and Gaussian pulses, the axial extent of the beams can be compared since all the depth-of-focus (DoF) of a Bessel beam carries laterally distributed rings that converge onto the optical axis. The same focal spot for a Gaussian beam can be achieved with focusing using $NA_G$ lens defined by $d^B_{min} = 1.22\lambda/NA_G$. Then, the DoF or double the Rayleigh length $2z_R = 2\pi[d^B_{min}]^2/\lambda = 18.8~\mu$m. This is $\sim 21$ times shorter than the DoF of the Bessel beam. Hence, the corresponding fluence of the Bessel beam reduced by this factor is 0.17~J/cm$^2$ and is close to the ablation threshold of Si.  
For the Gaussian pulse, the ablation threshold of Si is 0.2~J/cm$^2$~\cite{Jorn}.

\section{Results}

For this study, we chose \ce{Al2O3} mask due to its mechanical strength even at nano-film conditions and being compatible with anti-reflection coating of Si solar cells. Metal masks such as Cr or Ti make good adhesion to Si and can withstand plasma etch used in this study~\cite{22oea210086}, however, can deteriorate the performance of Si in terms of increased surface recombination velocity.

\subsection{Bessel beam patterning of \ce{Al2O3} mask}

The Direct Laser Write (DLW) by 515~nm/167~fs laser pulses with Bessel beam focusing was carried out on blank Si wafers coated by $30\pm 2$~nm and  $40\pm 2$~nm \ce{Al2O3} films deposited by electron (e)-beam evaporation. Single and double pulse per ablation site was used to write $1\times 1$~cm$^2$ masks at 7.5~cm/s scan speed; the laser repetition rate was 0.6~MHz. It takes $\sim 20$~min to pattern a $1\times 1$~cm$^2$ area. These scan and repetition rate conditions were chosen for the reliable positioning of ablation sites within $\pm 20$~nm tolerance from the designated positions on a square grid with $\Lambda = 3.1$~\si{\micro\meter} period as required for the PhC light trapping. The writing speed can be further increased since $\sim 30$~nm positioning errors were observed for the maximum tested writing speed of 10~cm/s; 
further increae of a writing speed will be studied separately. The positioning errors translate into mesa errors, especially at the corners of the etched pit. The orientation of the mask was aligned to the base cut of Si wafer to have the best etching conditions and smallest mesa regions. Irradiation of two pulses per ablation site caused a larger debris field as well as holes in the mask clogged by droplets of quenched melt (similar as in region-3 of Fig.~\ref{f-dumy}). It is noteworthy that debris on a mask and small changes in the shape of the hole-in-mask was still acceptable for plasma etch. Thinner 30~nm \ce{Al2O3} masks showed less debris and smaller holes, however, this thickness value was not selected for the final fabrication on Si solar cells due to being prone to fracture when undercut-etched. Thicker $\sim 50$~nm masks showed the presence of brittle fracture at the rim of the opening and larger debris fields.

\subsection{Plasma etch: optimization of conditions}

Firstly, different chemical mixtures for dry plasma etching were investigated with the aim of identifying those that yield the smoothest surfaces of Si production. One of the simplest mixtures of \ce{SF6}/\ce{O2} was the most promising since the addition of \ce{CF4} and/or \ce{CHF3} resulted in more nano-textured surfaces even if the etch rate was faster (for illustration see selected conditions shown in Fig.~\ref{f-chem}). 

The parameter space of optimal etch protocol was investigated with a change of Inductively Coupled Plasma (ICP) power, which favors isotropic all-direction material removal as well as Bias power, which defines anisotropic directional (between electrodes) etching. Figure~\ref{f-fast} shows optical images of pattern evolution at large ICP power with bias at 0~W. When the corner posts of Si below the mask become sub-1~\si{\micro\meter} in the lateral cross-section, they appear colorful under optical inspection with an objective with numerical aperture $NA = 0.9$ (see inset in Fig.~\ref{f-fast}(d)). Different colors are  related to the small differences in shape of pillars which evolves from small changes of ablated holes in the mask (see SEM inset image in (c)). This stage of color appearance in pattern formation is indicative that the plasma etch should be finished, since the \ce{Al2O3} mask breaks when the undercut etch progresses and some of the posts will not support the mask (38-nm-thick in Fig.~\ref{f-fast}). A mask is removed by immersion of the sample into an ultra-sonic bath of isopropyl alcohol (IPA) for five minutes. If the mask is not fully removed from the Si post, it appears colorful under optical inspection. Since alumina is part of the anti-reflection coating in the final design of the solar trapping surface, this is not considered a critical issue. For the final etch protocol, the used slow etch rate resulted in posts of the same color (dark-blue) under microscope observation, which indicates size and shape uniformity (Fig.~\ref{f-blue}).

The PhC pattern was defined by etch mask on a $1\times 1$~cm$^2$ area which had unablated $4\times 4$ holes every 1~mm in x and y-directions of the pattern by design. Those posts were added to support the \ce{Al2O3} membrane during dry plasma etch (this action is illustrated in Fig.~\ref{f-fast}(e) where mask is fixed to the post after it fails in the free-suspended region). Their contribution to an increased reflection from Si solar cell with PhC light trapping patterns can be considered small. 

Even though ICP is known for favoring isotropic etch, at lower ICP powers the etch becomes slower and the crystalline structure of Si becomes prominent on the etched surface (Fig.~\ref{f-topICP}). We chose low 30~W ICP power for the optimized protocol and no oxygen was used to reduce surface oxidation, which caused non-homogeneous islands/residue formation apparent under SEM inspection.  

Next, the influence of the bias power was investigated. Figure \ref{f-bias} shows side view cross sections aligned to the original surface of Si (a-c) etched under increasing bias, i.e. directional plasma bombardment of the surface. Elliptical markers are added as eye guides to facilitate a small increase of depth at large powers. The effect of bias was not very prominent understandably due to the small sub-1~\si{\micro\meter} opening in the mask through which the etching of Si was made. An increased bias power was also affecting the thickness of \ce{Al2O3} mask, which is undesirable. Zero bias was used for the final protocol. During SEM observation of samples after different times of plasma etch, it was observed that charging nano-debris are increasingly present (Fig.~\ref{f-bias}(c,d)) after repeated exposures of the sample to the air (in class 1000 cleanroom). This residue of etchants or/and oxidized Si is minimized with reduced steps of air exposures as in the final fabrication. With a lower content of oxygen in \ce{SF6}/\ce{O2} mixture, the surface had less of the oxidized residue and smoother surfaces. For the final protocol, no oxygen was selected. 

The process pressure in the chamber is another important parameter, especially due to the ICP-dominated etch, which was selected for the PhC pattern. The \red{He} gas flow below the sample was used for cooling the sample and was negligibly contributing to the process pressure. The reactivity of plasma and the isotropic nature of etching depended on the process pressure (Fig.~\ref{f-pres}). The low pressure facilitated the smooth morphology of Si with the largest depth (Fig.~\ref{f-pres}), which are both desirable virtues selected for the final protocol (shown in Fig.~\ref{f-pres}(a)). The sensitivity of etched surface roughness and presence of oxidized textures is illustrated in Fig.~\ref{f-topP} where ambient pressure changes are shown to strongly affect the morphology at larger ICP powers and \ce{O2} flow even at 0~W Bias.     

The key attribute of the proposed plasma etch protocol using a large area laser patterning of 40-nm-thick \ce{Al2O3} mask is timing. Etching should be stopped at the moment of the smallest pillar posts' formation (Fig.~\ref{f-under}). The posts provide mechanical support for the mask. Undercut etching and narrow mesa formation are already present even before posts are etched out. With the posts of $\sim 150\pm 50$~nm in cross-section, the deepest PhC pattern is formed with minimized mesa ridges. The final optimized recipe of PhC etch was selected for the slow etch, which caused the formation of the smallest posts (same deep blue color; Fig.~\ref{f-blue}), no oxygen in the etching gas mixture (only \ce{SF6}), 0~W Bias, 30~W ICP, and 1~Pa of process pressure. Such etching revealed the (111) Si planes typically observed in wet KOH etching. Minimum oxidation debris is formed on the surface, which is subsequently processed via wet etch KOH passivation and anti-reflection coating steps for the final solar cell. 

Figure~\ref{f-after}(a) shows the appearance of etched PhC pattern using \ce{Al2O3} mask patterned by fs-Bessel beam on the surface of actual solar cells. The total etch time to achieve minimal mesa ridges (Fig.~\ref{f-after}(b)) was approximately twice as long for the entire Si wafer as compared with the test samples of $\sim 1\times 1$~cm$^2$ patterned masks used for development of the etch protocol. However, the quality of the etched surface, mechanical integrity of the mask, and residue formation were the same as per the optimized recipe. One important difference of the wafer scale etch was the slight radial variation of undercut etching, which is an important and favorable attribute of the plasma processing of PhC light trapping surface (Fig.~\ref{f-after}). Dedicated wafer-size optimization of the etch protocol is required. For that, we foresee the requirement of a mechanically stronger mask for longer etching with an undercut etch conditions as discussed in the next section.       

\section{Discussion}

Figures~\ref{f-after} and \ref{f-2Rob} show currently achieved patterning of Polysilicon on Oxide  Interdigitated Back Contacted - POLO$^2$-IBC cell. 
This POLO$^2$-IBC architecture, same materials, and fabrication protocols achieved 26.1\% solar-to-electric power conversion efficiency~\cite{HAASE2018184}.
The mask writing throughput by the Bessel beam was 1~cm$^2$ in 20~min. A further increase is possible by combining xy-stage scan with glavano-scanning~\cite{19oe15205}. For the Bessel beam with the axial extent of the $\sim 400~\mu$m used in this study, there is a high tolerance for axial position of the focal region on the Si surface with 40-nm-thick mask even for an existing tilt of the sample and surface unevenness. Curved surfaces can be, in principle, patterned with the Bessel beam. In this study, the pulse duration corresponds to 50~$\mu$m in air as it travels through the non-diffracting section of $400~\mu$m. 

What was observed in the full wafer etch (Fig.~\ref{f-after}), is a twice longer etch time required for the same depth of the PhC pattern. This longer etch puts more stringent requirements against mechanical failure of the patterned mask. A higher ICP power would facilitate a faster etch rate, however, mask thinning with more erosive plasma would take place~\cite{23m550}. A practical solution to make better support for the etch mask would be to omit every second ablated hole in every second line (see concept in Fig.~\ref{f-dream}(a)). Finite difference time domain (FDTD) modeling (Lumerical, Ansys) of a toy-model of such PhC texture for 3-$\mu$m-thick Si is shown in (b) for the spectra of absorption and scattering cross-sections. Such a pattern has very similar cross-sections as compared with an inverted pyramid PhC. Figure~\ref{f-fdtd} shows light intensity distributions at very different wavelengths for a normally incident plane wave. The incident intensity is normalized and the $|E|$ shows the enhancement of electrical E-field due to interference inside Si especially at the longer wavelengths. This toy-model simulation qualitatively captures the wave nature of PhC light trapping which can perform above the Lambertian $4n^2$ ray optics trapping used in solar cells~\cite{SJohn2}.

The proposed model of the etch mask is amenable over large areas (10-inch wafer) and larger. Due to a mechanically stronger mask, it can potentially be thinner and smaller ablation holes can be printed if required for a smaller PC period. The very same etch chemistry at a higher ICP and bias were also tested for thinning of Si from 0.5 to 0.1~mm  (Fig.~\ref{f-thin}). Such thinning opens the possibility to use a compatible process to thin Si solar cells with backside contacts to the required thickness before adding the PhC trapping surface, which is the most efficient in thin tens-of-micrometers Si solar cells~\cite{SJohn1}.     

\section{Conclusions and outlook}

Direct laser writing of a hard etch mask using ultra-short fs-laser pulses is demonstrated for fabrication of PhC light trapping surface on actual high-efficiency Polysilicon on Oxide  Interdigitated Back Contacted - POLO$^2$-IBC cells. By using tens-of-nm \ce{Al2O3} hard mask ablated by fs-laser Bessel beam at high $7.5$~cm/s scan speed at room conditions, makes this approach scalable for solar cell applications. Periodic PhC patterns are required to break the ray optics Lambertian limit of light trapping. A detailed account of explored parameter space to optimize plasma etching is presented for different etch chemistries, powers of ICP and Bias, and process pressure. Only an optical inspection of the evolution of PhC formation is required for the determination of etching time and degree of under-etching. 

Scalability of such lithography-less and vacuum-less micro-patterning of PhC textures over even larger areas is outlined and  feasibility of light trapping is qualitatively evaluated using a numerical toy model. Detailed performance of these PhC-POLO$^2$-IBC cells will be reported in a dedicated study. Here, the experimental account for the selection of hard-mask thickness for direct laser writing by Bessel beam and dry plasma etching were all synergetically optimized and solved for an up-scalable application. 
 
\small\begin{acknowledgments}
We wish to acknowledge the support via ARC DP190103284 "Photonic crystals: the key to breaking the silicon-solar cell efficiency barrier" project (S.Ju., S.Jo.). 
This work was performed in part at the Melbourne Centre for Nanofabrication (MCN) in the Victorian Node of the Australian National Fabrication Facility (ANFF).
Support by the German Federal Ministry for Economic Affairs and Climate Protection, research project “27Plus6” (FKZ03EE1056A) is acknowledged. 
\end{acknowledgments}
\bibliography{
paper6}
\afterpage{\clearpage}
\appendix
\setcounter{figure}{0}
\makeatletter 
\renewcommand{\thefigure}{A\arabic{figure}}
\section{Appendix}

Here, a set of supplementary figures illustrates and summarises experimental observations with a selection of qualitative illustrations of the main consequence of particular etch parameter change. Figure captions are self-explanatory with technical details and main conclusions.     

\begin{figure*}[!h]
    \centering\includegraphics[width=18cm]{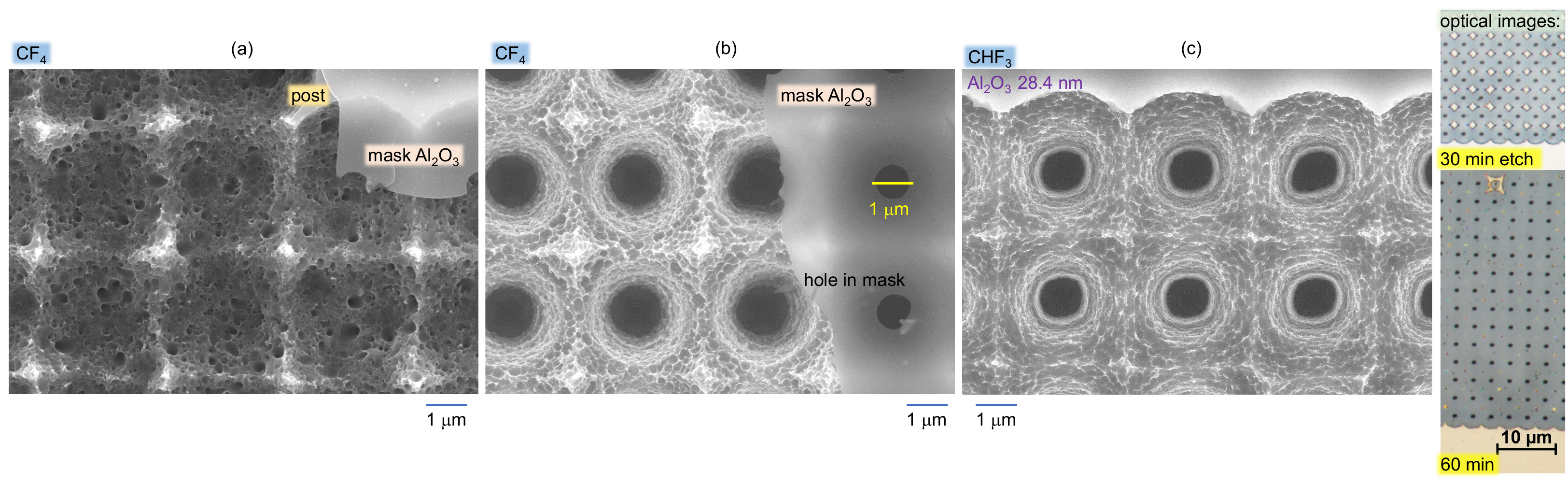}      \caption{\label{f-chem} Typical PhC structures at different etch chemistries. (a) \ce{CF4} test: 2~Pa process pressure, 200~W ICP, 0~W Bias, 50/5 ccm of \ce{CF4}/\ce{O2} for 60~min. Strong under-etch. Thickness of \ce{Al2O3} mask 37.6~nm, Bessel beam. (b)
same as (a) only 50~W ICP for 150~min. 
(c) Etch mixture with \ce{CHF3}: 2.5~Pa  process pressure , 180~W ICP, 0~W Bias, 50/10/10~ccm of \ce{SF6}/\ce{O2}/\ce{CHF3}. Thickness of \ce{Al2O3} mask 28.4~nm; Bessel beam was used for ablation. The inset in (c) shows optical images of etched pattern evolution, which turns into colored posts after 60~min of etch. 
}
\end{figure*}
\begin{figure*}[tb]
    \centering\includegraphics[width=14cm]{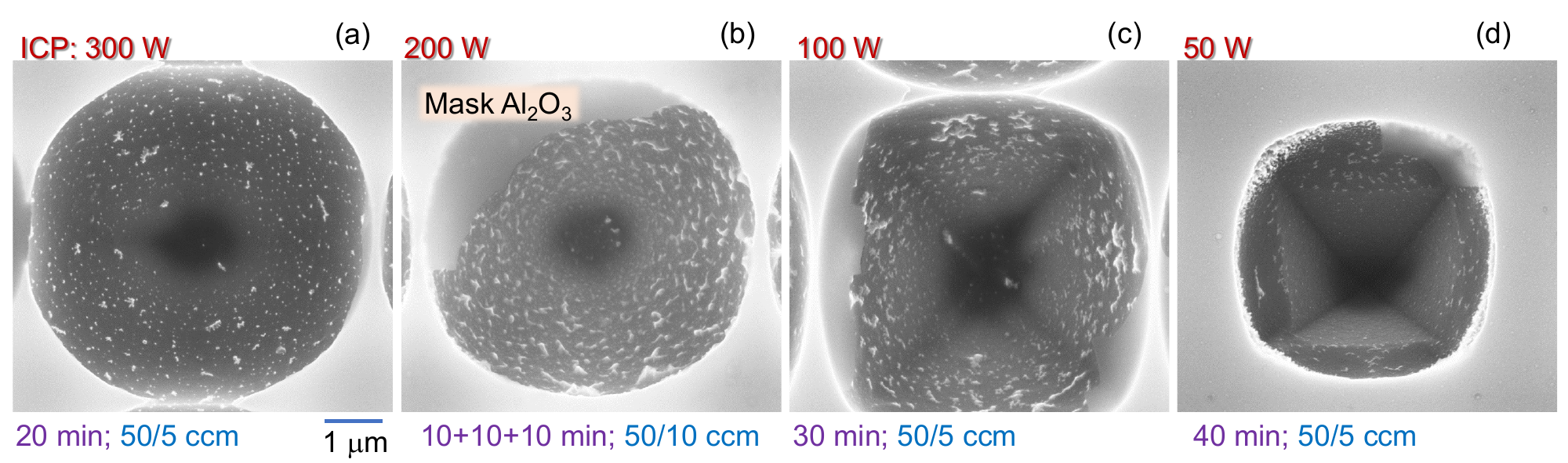} \caption{\label{f-topICP} Dependence of etch on ICP power in \ce{SF6}/\ce{O2} mixture (ratios are shown at the bottom of panels) at 0~W Bias. Representative SEM images of top-view after different times of etching. Presence of oxygen shows propensity of oxidation on the etched surface. At low ICP powers, crystolographic structure typical for the Si $\left<001\right>$ becomes discernable, see (a) vs. (d). The thickness of \ce{Al2O3} was 38~nm; mask writing by Bessel beam.
}
\end{figure*}
\begin{figure*}[tb]
    \centering\includegraphics[width=18cm]{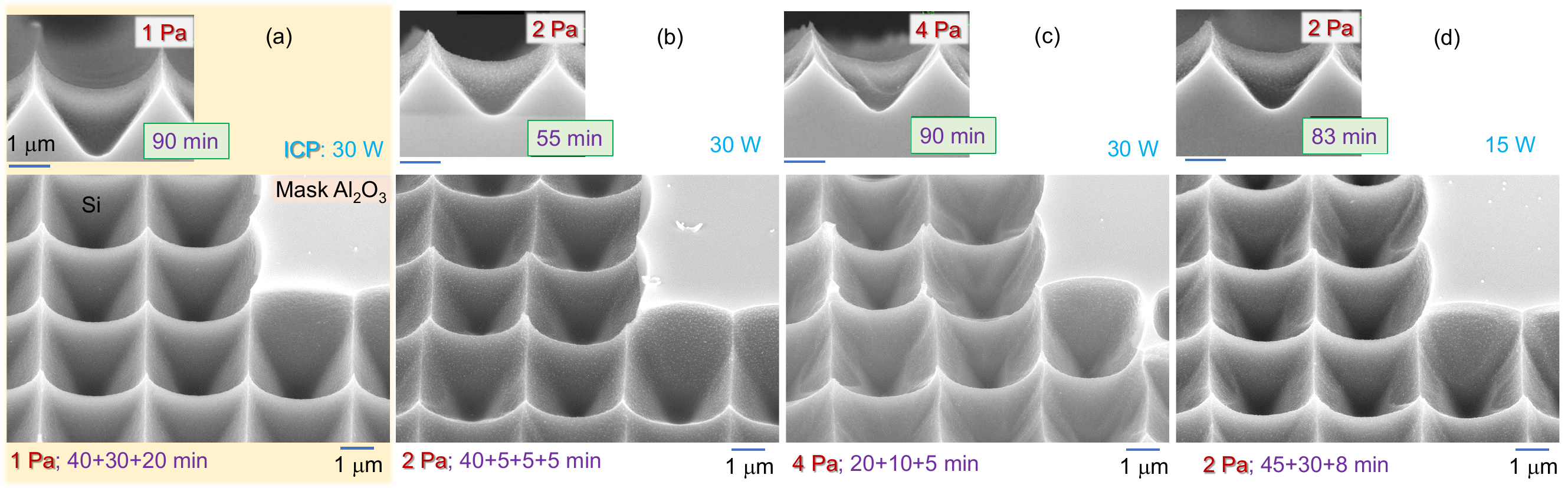}      \caption{\label{f-pres} SEM images of etched PhCs on Si at different process pressures: 1~Pa (a), 2~Pa (b), 4~Pa (c) and 2~Pa (d). Top-insets show side-view cross-section. Etch time is shown at the bottom of panels; when repeated etching was applied, times were shown with a summation marker ``+'' at the bottom of panels. RIE was carried out at 0~W Bias, 30~W ICP power (a-c) and 15~W (d), at 50/0~ccm \ce{SF6}/\ce{O2} gas mixture. Thickness of \ce{Al2O3} mask was 28.4~nm; mask writing was with Bessel beam. Conditions in (a) were selected for fabrication of PhC light trapping patterns on solar cells.
}
\end{figure*}
\begin{figure*}[tb]
    \centering\includegraphics[width=18cm]{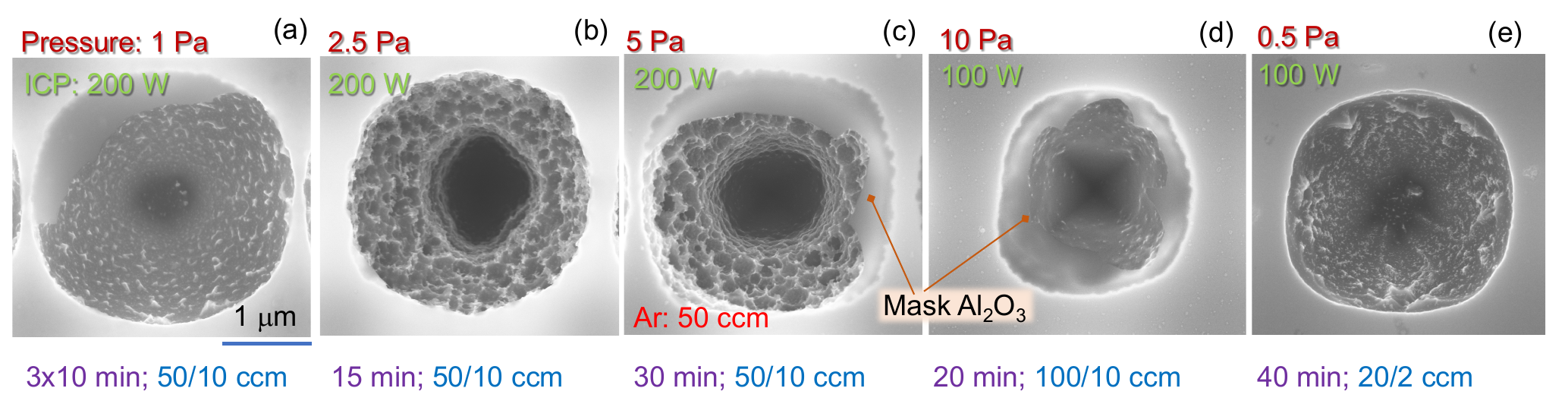}      \caption{\label{f-topP} SEM images of etched PhCs on Si at different process pressures: 1~Pa (a), 2.5~Pa (b), 5~Pa (c) and 10~Pa (d) and 0.5~Pa (e). Etch time is shown at the bottom of panels. RIE was carried out at 0~W Bias in all cases, while ICP was 200~W and 100~W (marked); Ar at 50~ccm was added for (c). The ratio of flow rate of \ce{SF6}/\ce{O2} gas mixture is shown below the panels. Thickness of \ce{Al2O3} mask was 28.4~nm; mask writing was with Bessel beam.
}
\end{figure*}
\begin{figure*}[tb]
    \centering\includegraphics[width=14.5cm]{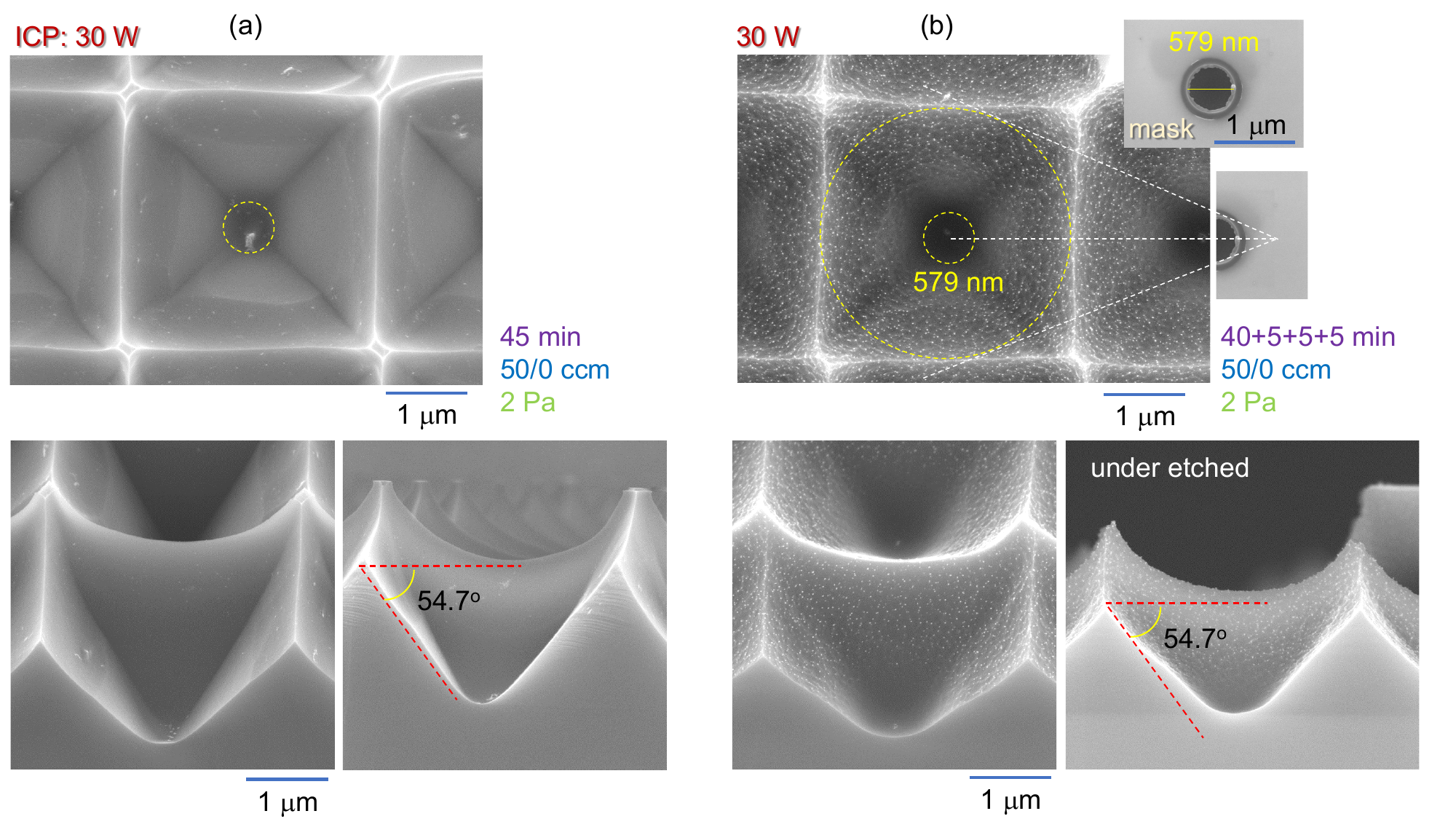}      \caption{\label{f-rep} Time, the ratio of flow rate of \ce{SF6}/\ce{O2} gas mixture, and pressure are show. Bias power was 0~W and that of ICP 30~W. Single run of etching (a) and with three repeated inspections (b) to achieve minimal mesa regions and under etching. Top-inset in (b) shows the typical opening in the mask used for this etching with its projections on the etched inverted pyramids as central circles. Thickness of \ce{Al2O3} mask was 38~nm; mask writing was with Bessel beam. The angle of $54.7^\circ$ between top surface (100) plane and (111) plane is shown. Repeated exposure to room conditions caused a larger amount of the oxidised residue shown in (b). Under etching conditions (b) affected the depth and slope of side walls and was stronger at larger ambient pressure (see Fig.~\ref{f-pres}).  
}
\end{figure*}
\begin{figure*}[tb]
    \centering\includegraphics[width=13.5cm]{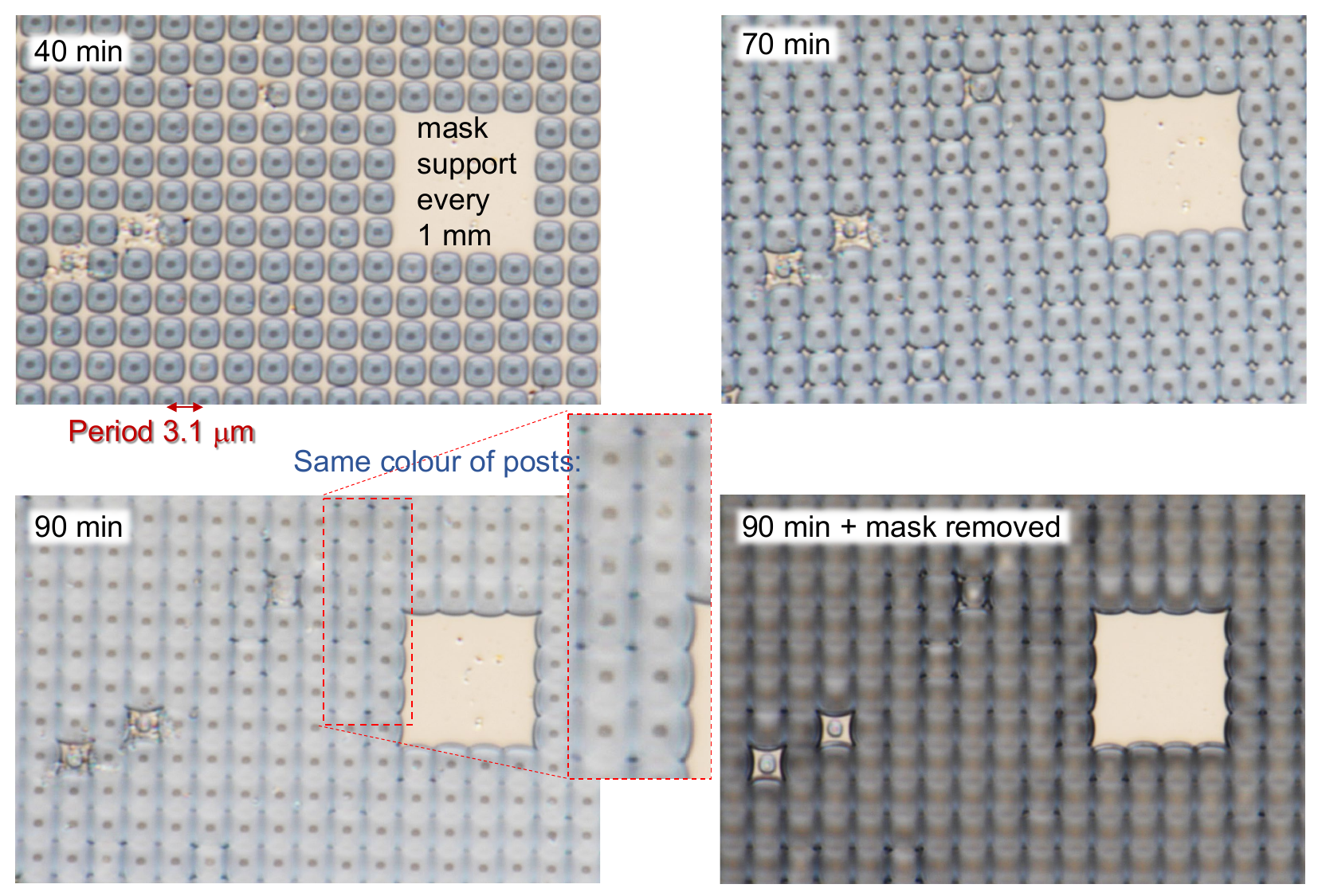}      \caption{\label{f-blue}  The optimal etch conditions. Optical images of etching progression for a pressure of 1~Pa, 30~W ICP, 0~W Bias, and 50/0~ccm of \ce{SF6}/\ce{O2} at the same position on the sample. The smallest corner posts are of the same colour (dark blue) observed after 90~min etch; see different colors at conditions of faster etch in Fig.~\ref{f-fast}(d). At this condition, etching is stopped and the mask is removed in ultra-sonic bath.  Thickness of \ce{Al2O3} mask was 38~nm; mask writing was with Bessel beam.
}
\end{figure*}
\begin{figure*}[tb]
    \centering\includegraphics[width=14cm]{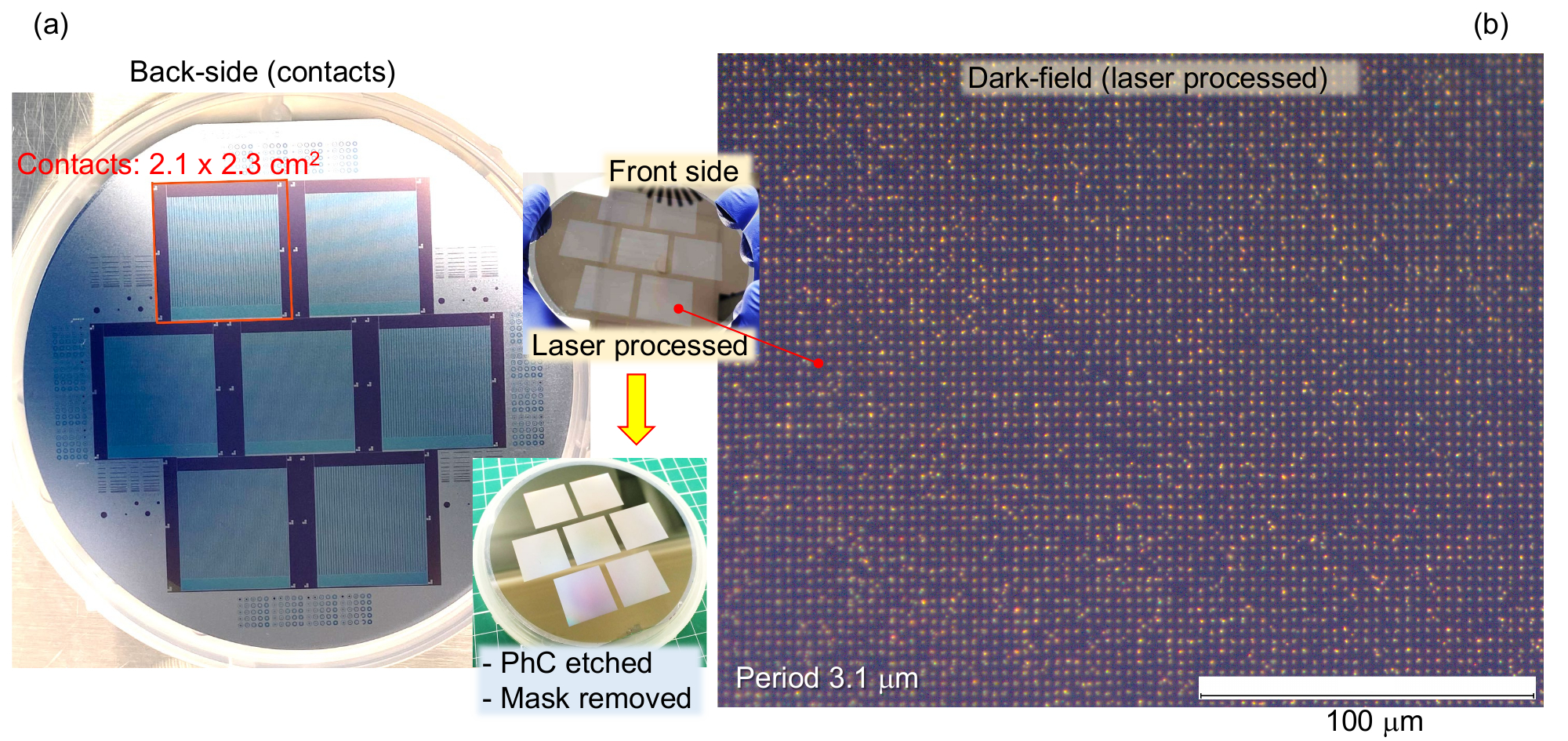}      \caption{\label{f-2Rob} (a) Back-side view of Polysilicon on Oxide  Interdigitated Back Contacted - POLO$^2$-IBC cell 
    with the top-inset image of patterned etch mask by Bessel beam 515~nm/167~fs, pulse energy $E_p = 43.8$~nJ. The thickness of Si solar cell 190~$\mu$m. Insets show laser patterned mask and final PhC light trapping surface and optimised plasma etch and mask removal. (b) Optical dark-field image of laser inscribed mask in 38-nm-thick \ce{Al2O3}.
}
\end{figure*}
\begin{figure}[tb]
    \centering\includegraphics[width=8cm]{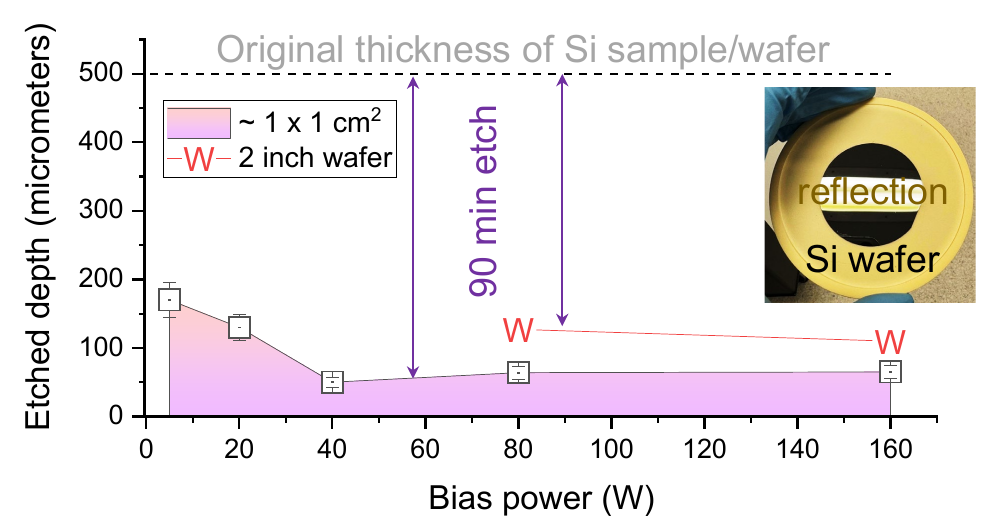}      \caption{\label{f-thin} Thinning of $500~\mu$m wafers down to sub-100~$\mu$m thickness (Samco ICP-RIE). Plasma thinning of 
    p-type Si at different bias powers, when ICP power was 700~W, process pressure 1~Pa, He back pressure 2~kPa, etching chemistry \ce{SF6}. Etching time was 90~min; error bars $15\%$. There was a strong dependence on the size of Si substrate, i.e. a  cm-sized sample was etched $\sim 50\%$ faster than a wafer. The most clean debris-free surface was obtained at the highest bias power of 160~W, when the etch rate was $4.6~\mu$m/min and the roughness was $\sim 25$~nm. Noteworthy, n-type Si had a strongly light scattering appearance at the visible spectral range after etching; the intrinsic and p-type Si was etched identically.  
}
\end{figure}

\end{document}